\documentclass{eas}
\usepackage{graphicx}
\usepackage{natbib}

\begin{document}

\title{Star formation in galaxy interactions and mergers} 
\author{Fr\'ed\'eric Bournaud}\address{CEA Saclay, IRFU, SAp, F-91191 Gif-sur-Yvette, France.}
\begin{abstract}
This lecture reviews fundamental physical processes on star formation in galaxy interactions and mergers. Interactions and mergers often drive intense starbursts, but the link between interstellar gas physics, large scale interactions, and active star formation is complex and not fully understood yet. We show that two processes can drive starbursts: radial inflows of gas can fuel nuclear starbursts, triggered gas turbulence and fragmentation can drive more extended starbursts in massive star clusters with high fractions of dense gas. Both modes are certainly required to account for the observed properties of starbursting mergers. A particular consequence is that star formation scaling laws are not universal, but vary from quiescent disks to starbursting mergers. High-resolution hydrodynamic simulations are used to illustrate the lectures.
\end{abstract}
\maketitle

\section{Introduction}
This lecture reviews the physical processes of Interstellar Medium (ISM) dynamics and Star Formation (SF) in galaxy interactions and mergers. Galaxy collisions and mergers are one of the fundamental processes of galaxy assembly in $\Lambda$-CDM cosmology. Mergers are often associated to "starbursts" where the rate of star formation can increase by a larger factor, and the gas consumption timescale drops. Here we aim at explaining the observed properties of these merger-driven starbursts with a theoretical point of view and with the help of hydrodynamic simulations, and compare to isolated disk galaxies that form stars quiescently. 

The fundamental processes of star formation in any galaxy are firs reviewed in a brief and simple way -- a thorough review of these is provided by Elmegreen (this volume; arxiv:1101.3108 to 1101.3113). The role of ISM turbulence is highlighted. We then study the dynamics of galaxy interactions and mergers, and the general properties of SF in mergers. The standard theory for merger-induced starbursts, based on tidally-driven gas inflows triggering a nuclear starbursts, is reviewed and its limitations are pointed out -- real starbursts tend to be stronger, and more extended spatially.  We show with modern high-resolution simulations that another triggering process is the increase of ISM turbulence and fragmentation, resulting from the interactions and inducing an excess of very dense, rapidly star-forming molecular gas. Merger-induced starbursts then have a nuclear component, but also an extended component in massive and dense star clusters, as generally observed. The gas compression is thus both global (over the whole galaxy) from the tidal field, and local (on scales of 0.1-1~kpc) from local shocks in the turbulence ISM. This explains recent observations, such as a dual law of star formation between quiescent disks and starbursting mergers, and an excess of very dense molecular gas in Ultra Luminous Infrared galaxies. 

Structure formation and (globular) star cluster formation/evolution, also included in this lecture, are briefly summarized at the end of the present text, and the reader is referred to \citet{bournaud-clusters} for more details on these aspects.

\section{Galaxy-scaled Star Formation and ISM dynamics -- basic facts}

\subsection{Disk stability and star formation}

Star formation on large scales proceeds by gravitational instabilities, which create dense clouds in which the gas cools more rapidly, and continue to fragment and collapse into start through gravitational (and thermal) instabilities. 

\begin{figure}[!ht]
\centering \includegraphics[width=3.5cm]{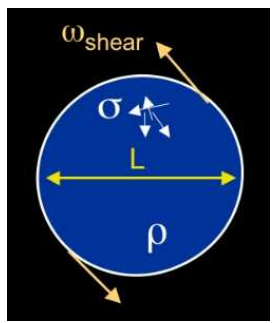}
\caption{An homogeneous region of volume density $\rho$ is unstable if its size is at least the Jeans length, which enables gravity to overcome pressure forces from the internal velocity dispersion $\sigma$. If the region is embedded in a rotation disk, the collapse can occur only if the size is small enough compared to the imposed shear ; the $Q<1$ Toomre criterium indicates that the combination of both constraints can be satisfied.}\label{fb1}
\end{figure}

The simplest prescriptions on whether a large (kpc-size) region of a galaxy will form stars and at which rate can be obtained by studying its gravitational stability. 
For a gas medium of size $L$, density $\rho$ and velocity dispersion\footnote{ This velocity dispersion can be thermal (microscopic) and/or turbulent (macroscopic).} $\sigma$, pressure forces on the surface vary as $L^2$, and necessarily overcome self-gravity, which scales as the mass, hence as $L^3$. For large-enough scale, gravitational forces eventually dominate and large-enough regions are gravitationally unstable (Fig.~\ref{fb1}). 

The critical scale, which is also the most unstable one (the fastest growing mode) is the Jeans scale, set by:

\begin{equation}
L_{\rm Jeans} \simeq \sqrt{\frac{\sigma^2 \pi}{G \rho}}
\end{equation}

If gravitational collapse occurs, gas becomes denser, so the cooling time drops and the gas effectively gets colder, making pressure forces more and more negligible compared to gravity. It is hence realistic to consider the collapse of gas clouds as a ``free fall'' process, i.e. not regulated by pressure forces, at least until the scales of magnetic pressure in dense cores are reached, or until star formation has taken place and young stars start to heat and ionize the gas. The gravitational free-fall time is given by:

\begin{equation}
t_{ff}=\sqrt{\frac{3 \pi}{32 G \rho}}
\end{equation}

At fixed density $\rho$ in an infinitely large system and without external rotation, a large-enough region will always be unstable. But the question to decide on dense gas cloud formation and subsequent star formation is whether sufficient mass can be gathered over a size not larger than the entire galaxy, and even over a size not larger than the disk scale height (above which the mass stops growing as $\propto \rho L^3$). Thus a given region of a galaxy may be dense enough or not to be gravitationally unstable and form dense gas and stars.

Another aspect is the stabilization by rotation and shear in galactic disks. Too large regions will be torn apart by the shear faster than they tend too collapse, namely faster than the gravitational free-fall time. This process stabilizes regions that are large enough to undergo large differential rotation. This is formalized by the \citet{toomre64} parameter:

\begin{equation}
Q = \frac{\kappa \sigma}{\pi G \Sigma}
\end{equation}

where $\Sigma$ is the surface density of the disk, $\kappa$ is the so-called epicyclic frequency, without a factor 2 from the angular speed $\Omega$ in galactic disks with standard rotation curves (3.36 replaces $\pi$ in the definition of $Q$ for a collisionless fluid, i.e. a stellar disk).

$Q>1$ means that any scale is stabilized either by pressure or rotation -- scales smaller than the Jeans length are stabilized by the gas pressure, and the Jeans length is already large enough to be stabilized by rotation.

Regions with $Q<1$ can collapse, fragment, form cold gas clouds and star-forming cores therein, and this collapse should take a gravitational free-fall time.

\subsection{The heterogeneous and turbulent ISM}

The simple description above could give a simple model to describe star formation in galaxies in theory and in simulations. Only regions where $Q<1$ can form stars, and then a certain fraction of gas is converted into stars within one or a few free-fall times, the involved fraction being much smaller than one (typically a few percent) because the formation of the first stars will heat the surrounding gas and regulate the formation of the next stars (although young stars can also remotely trigger the formation of other stars \citep[e.g.,][]{elm-lada}).

However, the involved critical scale in this "$Q$$<$1 approach", the Jeans length or Toomre length (this two scalelengths are of the same order of value), is of the order of 100-1000~pc for normal spiral galaxies such as the Milky Way. If we consider a portion of the ISM of a few hundreds of pc, we will find warm atomic gas at densities of a few atoms per cubic centimeter and temperatures of a few $10^3$~K, colder and denser atomic gas, molecular gas with densities of $10^{3}$ to $10^6$~cm$^{-3}$, ionized gas at $\sim 1$~cm$^{-3}$ and a few $10^4$~K, etc. It is then impossible to define a typical density and velocity dispersion (or sound speed) in the ISM on the scales that are relevant to measure a Toomre parameter, making this theory impossible to apply to real galaxies.

The only framework in which the basic theory of gravitational instability in disks can be directly applied are maybe the resolution-limited SPH simulations that are a plenty in the literature since almost two decades. Most of these simulations (except a few recent ones) do not resolve an heterogeneous ISM, the only density fluctuations arising on scales of a kpc and above. There a relevant Toomre parameter can be measured and applied as a star formation criterium. But at the same time, this simple fact means that the involved ISM model has a quite limited realism.

\medskip

If we consider the various phases of the ISM in the Galaxy, up to one third of the volume may be filled with warm and hot gas above $10^4$~K, but at least two thirds of the mass (and 100\% of the mass involved in dense gas cloud formation and star formation) are in denser and colder phases with temperatures between, say, 10~K and $10^{3-4}$~K. The sound speed in atomic hydrogen at $10^4$~K is about 10~km~s$^{-1}$, thus the star-forming phases of the ISM have a sound speed of the order of a km~s$^{-1}$ (0.1 to 10~km~s$^{-1}$).

Observed linewidths of the gas in spiral galaxies are at least a few km~s${^1}$, they can reach 50 or 100 km~s${^1}$ in interacting galaxies (see details and references later), starbursting galaxies, and in gas-rich galaxies at high redshift. This is observed even on scales much smaller than the rotational velocity gradient in disk galaxies, which shows that supersonic non-circular motions are ubiquitous in the ISM of galaxies. 

Interestingly, observations of the density power spectrum of nearby galaxies suggest that these random motions follow a cascade process from a large injection scale down to dissipation on small scales \citep{block, dutta, E93}. The spectral signature of this cascade is just as expected from turbulence experiments \citep{petersen}. This indicates that the non-circular motions in the ISM can be named {\em turbulence}. A key property of this ISM turbulence is that it is {\em supersonic}, since the involved motions largely exceed the sound speed. Mach numbers of a few for the cold star-forming ISM phases are typical for spiral galaxies, they may frequently reach $\sim 10$ in starbursts, and high-redshift galaxies \citep[e.g.,][]{FS09}. 

\begin{figure}
\centering \includegraphics[width=7cm]{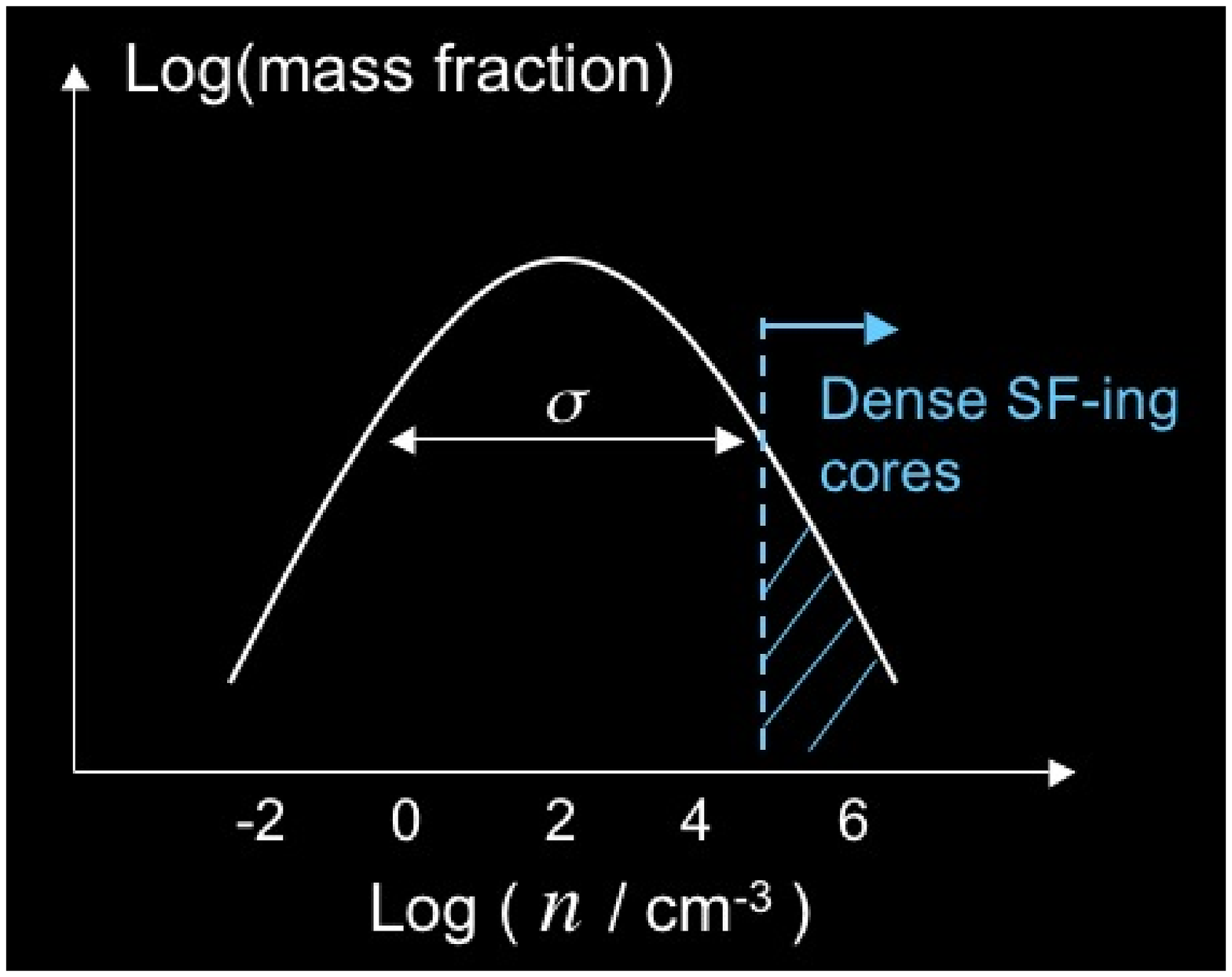}
\caption{This figure illustrates a density PDF, with a log-normal shape and typical value for the galactic ISM. The intrinsic spread of the PDF increases with the Mach number of the non-circular (turbulent) motions in the ISM \citep{wada02,K07}. The log-normal shape exist is the turbulence is (1) isothermal (but in the real non-isothermal ISM the high Mach numbers keep the PDF quasi log-normal anyways) in a steady-state (i.e. for an isolated self-regulated disk but not necessarily for a suddenly disturbed system in an interaction).}\label{fb2}
\end{figure}

\subsection{Star formation in the turbulent ISM}

The connection between ISM turbulence, dense gas phases and star formation is best understood by examining the point distribution function of the local gas density (density PDF). In its mass-weighted (resp. volume-weighted) version, the density PDF represents the mass (volume) fraction of the ISM in bins of density. We here use mass-weighted versions. 

In an system at equilibrium (e.g. isolated disk galaxy), supersonic turbulence generates a log-normal density PDF \citep{wada02} (see Fig.~2). Only small mass fractions of the whole ISM are at found number densities below $\sim 1$~cm$^{-3}$ or in very high density regions above, say, $10^4$~cm$^{-3}$, most mass being at ten to a few hundreds of atoms per cm$^{3}$ (typically observed as HI or CO-traced molecular gas). The width of the log-normal PDF depends on various factors, but the main dependence is on the turbulent Mach number, with the PDF width $\sigma$ increasing when the Mach number increases \citep[see][for details]{K07}. A more turbulent ISM will have a larger spread in its density PDF, because the turbulent flows will sometimes diverge and create low-density holes, and sometimes converge (or even shock) into very dense structures that can further collapse gravitationally. Note that log-normal density PDFs are observed, at least at the scale of molecular complexes \citep{alves}. 

The density PDF is a useful tool to describe the star-forming activity of a galaxy at a given gas content. For a simplified description, one can consider that star formation takes place only in the densest gas phases (i. e. above some density threshold), and that in these dense regions the local star formation rate follows, for instance, a fixed efficiency per free-fall time: $\rho_{\mathrm{SFR}} = \epsilon_{ff} \rho_{\mathrm{gas}} / t_{ff} \propto \rho_{\mathrm{gas}}^{1.5}$ (see detailed theory in \citealt{E02} and \citealt{K07}). Even if the local rate of star formation follows a different behavior than this purely density-dependent model (which is physically motivated by the gravitational collapse timescale), then the first step remains that star formation proceeds only in the densest gas phases once dense-enough clouds have formed. The weight of this "star-forming phase" is given by the high-density tail of the PDF. Hence, the fraction of dense gas along the density PDF is a key parameter for the global star formation activity of any given galaxy. The density PDF will thus be a major tool to understand starburst activity in galaxy mergers.


\section{Dynamics of galaxy interactions and mergers}

\subsection{Dynamical Friction}

When two galaxies start orbiting around each other, the first long-range process that drives their evolution is dynamical friction. When one object (e.g., the stellar body of a galaxy) flies through another mass distribution (e.g., the dark matter halo of the other galaxy), the gravitational forces will unavoidably create a mass accumulation from the latter system along the wake of the first object. This will in turn induce braking forces. This process also acts on non-overlapping systems, although it is much weaker then. Dynamical friction is what drives the merger of interacting galaxies into single, more massive objects. In the process a small fraction of the mass is expulsed and becomes unbound, carrying away a substantial fraction of the initial energy at high velocity, so that the merger remnant is a lower-energy, usually more compact system. 


\subsection{Violent Relaxation}

The typical merger timescale for a pair of massive galaxies is about a billion year, although large variations are possible for small impact parameters or high encounter velocities. This is barely larger than the rotation period of the outer disk of a spiral galaxy. Thus, stars (or gas clouds or dark matter particles) in an interacting and merging system undergo a variation in the mass distribution/gravitational potential that is relatively rapid compared to their own dynamical timescale. In such conditions stellar orbits do not follow a slow secular evolution anymore, but can be completely redistributed. For a given star, the average energy along its final post-merger orbit can be quite different from the initial average energy along the initial pre-merger orbit: this large difference arise because of the rapid change in the mass/potential distribution and defines the {\em violent} relaxation. Some stars (or gas clouds or dark matter particles) will then undergo large decrease of their average energy, which implies that they sink towards the central regions: the central regions of galaxies thus become denser and more compact in a merger. Other stars will gain energy and increase their orbit semi-major axis -- some even escape: a more extended outer envelope in the mass distribution results. 
 
When one writes the projected radial surface density profile of a galaxy as a Sersic profile $\log(\Sigma) \sim \mu \sim e{^-(r-a)^{1/n}/b)}$, where $n$ is the Sersic index, spiral galaxies have exponential disks with $n \simeq 1$. A merger remnant will have a steeper central profile and a more extended outer envelope, which corresponds to $n>1$. The high Sersic indices of elliptical galaxies (say, $2<n<6$) are thus consistent with the common idea that they are the end-product of spiral-spiral mergers.

\subsection{Gas response and tidal torques}
\paragraph{The tidal field}

A first driver of the gas response in galaxy interaction is the tidal field. Typically, the tidal field of a single galaxy is disruptive at large radius and outside its own disk, but can be compressive in the inner regions if the mass distribution has a low Sersic index. During a distant interaction or the early stages of a merger, a given galaxy is then mostly affected by a disruptive tidal field from its distant companion. More advanced mergers can have compressive tidal field over large regions, when the galaxies begin to overlap \citep[e.g.,][for the Antennae]{renaud}. 

While a disruptive tidal field naturally tends to expel material from disk galaxies, it is not the only mechanism responsible for the formation of long tidal tails, and it cannot explain central gas inflows: this is in fact mostly driven by gravity torques.

\begin{figure}
\centering \includegraphics[width=8cm]{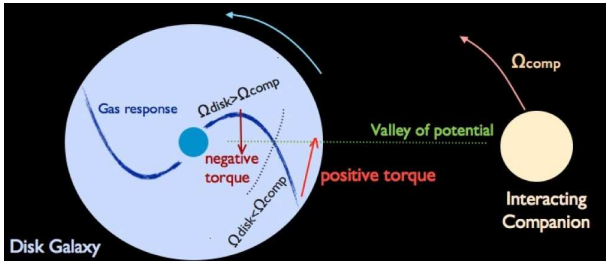}
\caption{Simplified description of the gas response in a galaxy interaction. Inside the corotation radius, gas concentrates on the leading side of the valley of potential and undergoes negative gravity torques, which drives a central gas inflow. At the opposite, positive gravity torques dominate outside the corotation, which drives lot of the outer disk material in the so-called ``tidal'' tails. In detail the gas response is more complex because the system evolves rapidly in a non-equilibrium configuration, but the main response is the one shown here. The same response occurs in barred galaxies in a weaker but more directly observable way (see Figure~\ref{fb5}).}\label{fb4}
\end{figure}

\paragraph{Gravity torques, inflows and outflows}
The tidal field from a companion breaks the symmetry of the gravitational potential. This induces a response of the disk material, in particular its cold gas, which is more pronounced for prograde\footnote{in which the orbit and disk rotations have similar orientations} interactions. The gas can form an interaction-driven pair of grand-design spiral arms, like modeled in detail for instance for M51 by \citet{dobbs51} -- note though that grand-designs spiral do not require an interaction to form \citep{ET93}, and are not connected to interactions in observations \citep{vandenbergh}. The gas response is often more complex than a pair of spiral arms, but a general feature remains: inside the corotation radius\footnote{The corotation radius is the radius at which the rotation speed of the studied disk equals the orbital speed of the companion.}, the gas preferentially concentrates on the leading side of the valley of potential; outside of the corotation it concentrates on the trailing side. A similar but better-known gas response is found in barred galaxies, where the disk symmetry is broken about the same way by the barred pattern \citep{CG85,athanassoula}. A corotation does not always exist and can move with time, but is in general located at a few kpc of radius in the disk. The subsequent process is illustrated on Figure~\ref{fb4}: inside the corotation, the gas undergoes negative gravity torques, and loose angular momentum in a rapid central gas inflow towards the central kpc or less. In the outer disk, gas would gain angular momentum and fly out to larger radii in long tidal tails.

\begin{figure}
\centering \includegraphics[width=8cm]{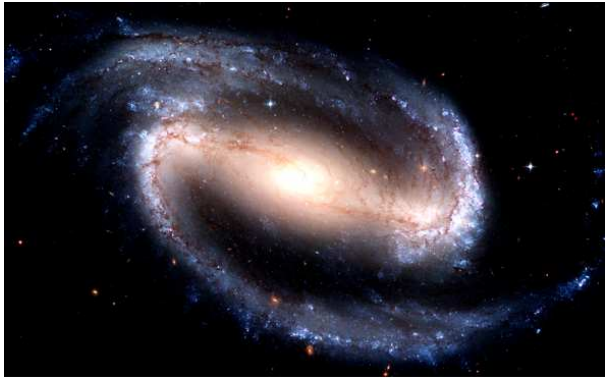}
\caption{In a barred galaxy, the valley of potential along the bar induces the same gas response as in an interaction (but at a weaker degree). The gaseous arms on the leading side of the bar inside the corotation radius, attested by two sharp dust lanes on this optical image, are more visible as the system evolves in a steady state.}\label{fb5}
\end{figure}

It is worth noting that:
\begin{itemize}
\item The so-called ''tidal'' tails do not just result directly from the tidal field. Gravity torques induced by the interaction are an important process in tidal tail formation. If the tidal field was the main factor, few mergers would actually have a central gas inflow and related central starburst. This was already illustrated by the restricted three-body simulations by \citet{toomre77}: the tidal field was included but not the full disk response, tidal tails were forming but relatively short and no central gas inflow was induced. The role of disk self-gravity and gravitational torques in amplifying the disk response was noted in \citet{toomre81}

\item Stars in a disk galaxy behave in a roughly similar way, but their much larger velocity dispersion and collisionless dynamics makes the gravity torques much weaker. This is why stars pre-existing to a galaxy interaction are generally not found in great amounts in tidal tails (e.g. Duc et al. 1997, but see for instance Duc et al. 2000 in Arp245). For the same reason, there is no significant central inflow of pre-existing stars (while the inflow of gas can form a lot of new stars in the center).
\end{itemize}
{\bf Collisional ring} galaxies form through a different dynamical process. The central impact of a companion onto the disk triggers large epicyclic motions around the initial orbits of stars and gas clouds. The superposition of motions and the variation of the epicyclic frequency with radius result in an expanding wave of material. This process is thoroughly reviewed by \citet{appleton}. The ring wave is initially an unbound feature, like the spiral waves in an isolated disk, but the ring can become self-gravitating, bound, when enough mass is gathered in the ring \citep[e.g.,][]{B07}. While the initial process differs, collisional rings and are relatively similar to tidal tails in terms of substructure formation, so they will hereafter be treated as the same kind of "tidal" structures.


\section{Star formation triggering in interactions -- From mergers to starbursts}

\subsection{Main observed facts}

\paragraph{Observations of a starburst--merger connection}
The main and simplest observation probing the link between starbursts and mergers is that almost all starbursting galaxies in the nearby Universe are mergers. In particular, (Ultra-)Luminous Infrared Galaxies (LIRGs and ULIRGs) at low redshift are almost exclusively strongly interacting and merging systems \citep[e.g.,][]{duc97}. Note that at high redshift ($z\sim 1$ and above), LIRGs are not necessarily mergers because ``normal'' isolated disk galaxies are gas-rich enough to reach LIRG-like activity \citep{daddi10a, tacconi10}, but it remains true that the strongest starbursts (ULIRGs and Hyper-LIRGs) are predominantly merging systems \citep[e.g.,][]{elbaz-cesarsky}. 

It has been long known that mergers {\em can } trigger bursts of star formation \citep{sanders88}. But the fact that the most actively star forming systems at any given redshift are mostly mergers (LIRGs at $z \sim 0$, ULIRGs at $z \sim 1$) does not necessarily imply that mergers are always starbursting, or that all mergers even undergo a significant starburst phase -- we here of course consider ``wet'' mergers where at least one galaxy contains a substantial amount of gas, not ``dry'' mergers of gas-free galaxies. Understanding whether all mergers are starbursting requires to be able to select mergers (at least major ones) in large surveys including star formation tracers. The main concern is that any technique selecting mergers based on morphology \citep[for instance those by][]{conselice-cas, lotz-gm20} remains ambiguous, in particular at high redshift where morphological disturbances and asymmetries can arise internally \citep{bournaud08}. Surveys including spatially-resolved gas kinematics are certainly a more robust tool to identify on-going major mergers and (relatively) isolated disks \citep{shapiro08} but such samples remain too limited to study the merger-starburst connection. Nevertheless, the most recent and detailed studies \citep{bell, jogee, robaina09} are broadly in agreement that mergers do trigger star formation activity, and specific star formation rate (i.e., SFR divided by the stellar mass already in place) being increased by factor of about 3-4, on average: strongly starbursting mergers do exist but most mergers do not increase the SFR by a factor above about 10 at the peak, likely because strong starbursts are rare, or they are short phases compared to the duration of an interaction/merger until final relaxation of the post-merger early-type galaxy. 

The factor 3-4 of increase in the SF activity is the average observed value for randomly interactions/mergers at any instant of the merging process, not at the peak of a starburst activity, so it likely indicates that peak SFR values are higher than this. Also, it is the typical measure for "major" interaction/mergers, including mass ratios of, say, 1:1 to about 3:1. There is no evidence of a variation with redshift, suggesting that the typical SF activity of ``isolated'' disks and merging systems increase with redshift in roughly similar proportions.

\paragraph{Morphology of merger-induced star formation}

Merger-induced star formation consists, for a large part, of nuclear starbursts taking place in the central 100-1000~pc (diameter). Nuclear activity in merger-induced starbursts has long been emphasized in studies of the merger-SF connection \citep[e.g.,][]{kennicutt87, sanders88,duc97, soifer84, lawrence89}. Nevertheless, the importance of nuclear starbursts among merger-induced star formation in general has long been overstated in the literature. This may be partly because theorists have long been able to explain only nuclear starbursts, or because of using infrared and/or H$\alpha$ luminosities to trace star formation -- these are dependent on metallicity, dust properties, and threshold effect making them more sensitive to compact central starbursts than to spatially extended star formation with lower surface density. 

Actually, there is a large fraction of systems in which merger-induced star formation is spatially extended, taking place outside the central kpc, and in these system the extended star forming component is not just a relics from extended SF in pre-merger spiral galaxies, but participates to the total starburst activity, even if there is also intense SF in the nucleus. The relatively large spatial extent of SF and the problem it poses for theoretical models (as we will review below) was probably first recognized by \citealt{barnes04} in the Mice (NGC~4676A+B, see also \citealt{chien-barnes10} in NGC7252). A well-know example if actually the Antennae system (NGC~4038+4039), where a large fraction of the starburst activity takes place at several kpc from the nuclei in big ``super star cluster'' (SSCs) formed along density waves throughout the interacting disks, and in the shocked overlap region between the two gas disks \citep{wang04} -- these star-forming components are more important than the dust-enshrouded nuclear star formation revealed in the infrared. There are many other examples of interacting/merging systems with spatially extended starbursts whose total SFR is far from being dominated by the nuclear SF activity, but rather by extended star-forming components: Apr~140 \citep{cullen06}; NGC2207+IC2163 \citep{ngcic-spitzer-e06} in which the starburst is almost exclusively non-nuclear with an SFR of 15~M$_{\odot}$~yr$^{-1}$ for global gas mass/size about similar to typical Milky Way-type spirals (i.e., a significant increase of SFR likely results from the interaction, but only in a spatially extended mode); the Antennae; the Cartwheel system -- see also many tidally interacting pairs in \citet{smith}, \citet{hancock}, etc. There is to my knowledge no thorough census of the contributions of nuclear and extended star formation to merger-induced activity, probably because unveiling dust-enshrouded nuclear SF and detecting spatially extended SF at lower surface densities requires to employ different techniques at different wavelength. Nevertheless, eye examination of systems with UV imaging available in samples of interactions and mergers (e.g., the Arp Atlas) suggest that spatially extended ($>$1kpc) SF and nuclear activity (within the central kpc) each contribute to a rough 50\% of star formation in mergers.

\paragraph{ISM properties in interacting starbursts}

Two fundamental properties of the interstellar gas in star-forming mergers majors will hold a fundamental role in our theoretical interpretation of the starburst activity:

\begin{itemize}

\item gas velocity dispersions are higher in interacting/merging systems than in isolated galaxies. At redshift zero, the cold ISM of spiral galaxies has a typical velocity dispersion of, say, 5--10~km~s$^{-1}$, and only gas-rich disk at high redshift have higher dispersions. Major mergers can have gas velocity dispersions of several tens of km~s$^{-1}$, as first observed by \citet{irwin} and \cite{E95} (see also examples in \citealt{bournaud-fp} although limited to star-forming clumps in H$\alpha$). These increased dispersions do not result from blending of several components in a large telescope beam (a.k.a. beam smearing), as they are also observed at high resolution, but trace a physical velocity dispersion on small scales. As this applies primarily to gas components observed through HI or CO lines, the thermal dispersion (sound speed) is lower than 10~km~s$^{-1}$ (for temperatures below $10^4$~K), so most of the observed dispersion in isolated galaxies and merging systems consists in supersonic gas turbulence, which usually dominates the cold star forming phases of the ISM \citep{burkert-ism} (a warm, thermally-supported phase can fill a large volume fraction, but involves only a minor mass fraction of the ISM). Hence, the (supersonic) ISM turbulence is typically increased by a factor of a few in galaxy interactions and mergers. We will discuss later, using numerical and theoretical models, that this is probably a trigger of increased star formation (through compressive regions in the turbulent flow) rather than a consequence of the star forming activity itself (through stellar feedback impacting the surrounding gas). 

\item Various density phases of the ISM are not filled and distributed in the same way as in isolated disk galaxies. There is often a spatial segregation between the moderate density, atomic phase, mostly found in the outer regions, and the denser molecular phases concentrated towards the few central kiloparsecs \citep[e.g.,][]{duc97}. More systematically, starbursting systems in the nearby Universe with ULIRG-like infrared luminosities (which are almost exclusively merger-induced starbursts, see above) exhibit enhanced HCN/CO luminosity ratios \citep{juneau09}. HCN is usually considered to be a tracer of the very dense gas (say, $\geq 10^{5-6}$~cm$^{-3}$) directly involved in actual star formation, while CO lines also (and mostly) trace lower-density regions (down to, say, 100~cm$^{-3}$) that belong to star-forming clouds but are not directly star forming: hence observations point towards an excess of high-density gas in starbursting mergers compared to the amount of moderate-density molecular gas traced by CO.

\end{itemize}

As detailed in the next Section, if the density PDF of the ISM is driven by the supersonic turbulent support, then the increased velocity dispersion observed in mergers can directly result in a larger spread of the density PDF, which can result in an excess of high-density molecules compared to the total molecular gas mass. We will illustrate later, using hydrodynamic simulations, how these processes act in concert to trigger the star formation activity of interacting and merging galaxies.

\subsection{Standard theory of merger-induced starbursts: gas inflows and nuclear starbursts}

In an interacting galaxy pair, the gas distribution becomes non-axisymmetric, which results in gravitational torquing of the gas as detailed previously. Gas initially inside the corotation radius (typically a radius of a few kpc) undergoes negative gravity torques and flows inwards in a more and more concentrated central component (inside the central kpc). Any model for star formation will then predict an increase of the star formation rate (global Schmidt-Kennicutt law, models based on cloud-cloud collisions, etc). The result is thus a centralized, nuclear starburst. As the driving process is gravitational torquing, early restricted three-body models could already describe the effect \citep{TT72}. Later models have added extra physical ingredients leading to more accurate predictions on the star formation activity: self-gravity \citep{BH92}, hydrodynamics and feedback processes \citep{mihos, cox-ratios}, etc.

A large library of SPH simulations of galaxy mergers, in which the driving process is mostly the one presented above (tidal torquing of gaseous disk) was performed and analyzed by \citep{dimatteo07,dimatteo08}. This study highlighted various statistical properties of merger-induced starbursts. In particular, it showed that some specific cases can lead to very strong starbursts with star formation rate (SFRs) are increased by factors of 10--100 or more, but that on average the enhancement of the SFR in a random galaxy collision is only a factor of a few (3--4 being the median factor) and only lasts 200-400~Myr.  These results were confirmed with code comparisons, and found to be independent of the adopted sub-grid model for star formation \citep{dimatteo08}. Models including an external tidal field to simulate the effect of a large group or cluster found that the merger-induced starbursts could be somewhat more efficient in such context -- but the SFR increase remains in general below a factor of 10 or less \citep{MB08}.

\subsection{Limitations of the standard theory}

{\bf \noindent The intensity of merger-induced starbursts}
Numerical simulations reproducing the interaction-induced inflow of gas and resulting nuclear starbursts can sometimes trigger very strong starbursts, but in general the SFR enhancement peaks at 3--4 times the sum of the SFRs of the two pre-merger galaxies. This factor of 3--4 seems in good agreement with the most recent observational estimates \citep[e.g.]{jogee, robaina09}. In fact, there is substantial disagreement: the factor 3--4 in simulation samples is the peak amplification of the SFR in equal-mass mergers. In observations, it is the average factor found at random (observed) instant of interactions, and in mergers that are "major" ones but not strictly equal-mass ones. Given that typical duration of a merger is at least twice longer than the starburst activity in the models, and that unequal-mass mergers make substantially weaker starbursts \citep{cox-ratios}, one would need a peak SFR enhancement factor of about 10--15 (as measured in simulations) to match the average enhancement of 3--4 found in observations. There is thus a substantial mismatch between the starbursting activity predicted by existing samples of galaxy mergers, and that observed in the real Universe -- although the observational estimates remain debated and may depend on redshift.
\medskip

\begin{figure}
\centering \includegraphics[width=6cm]{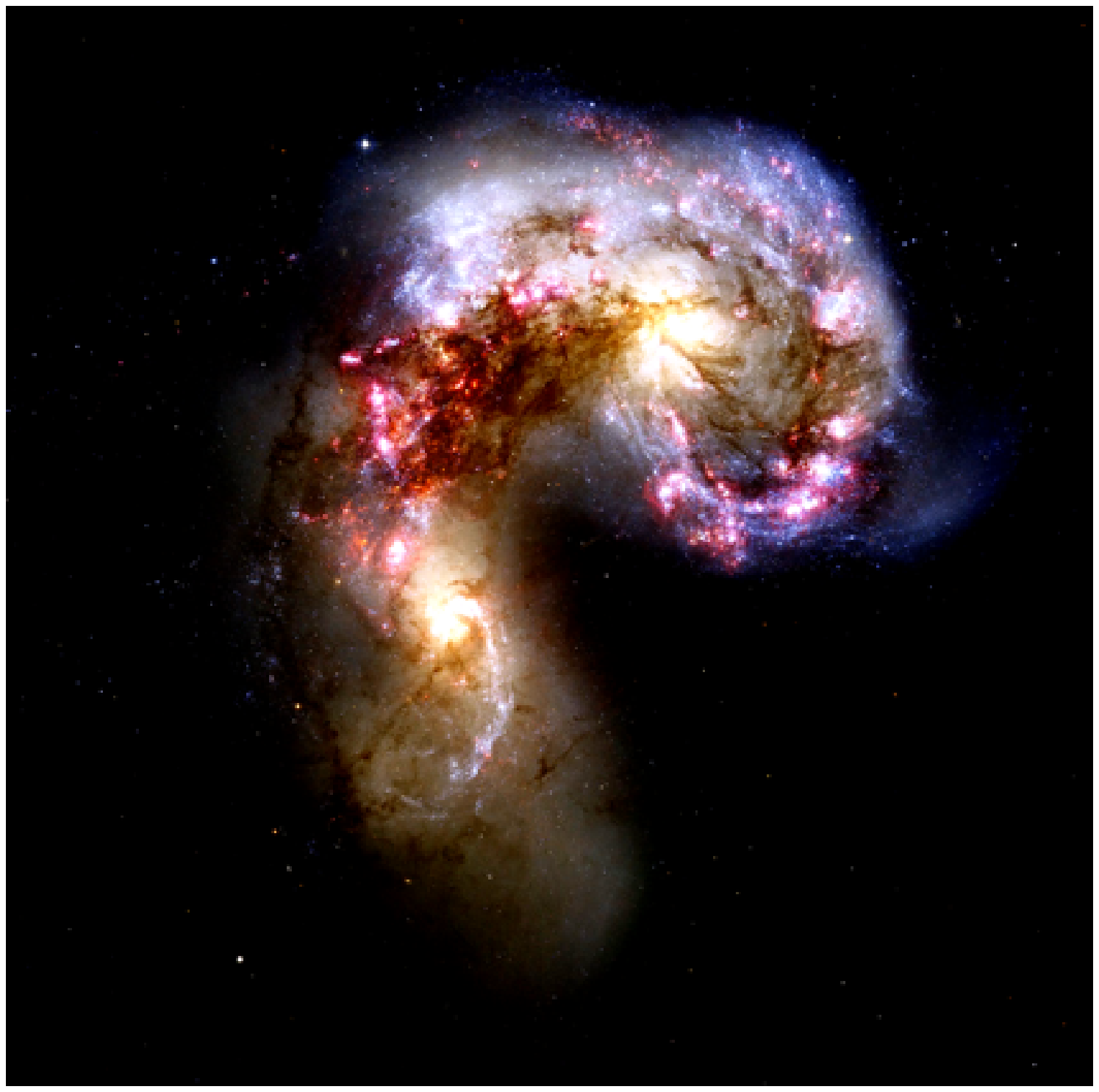}
 \includegraphics[width=6cm]{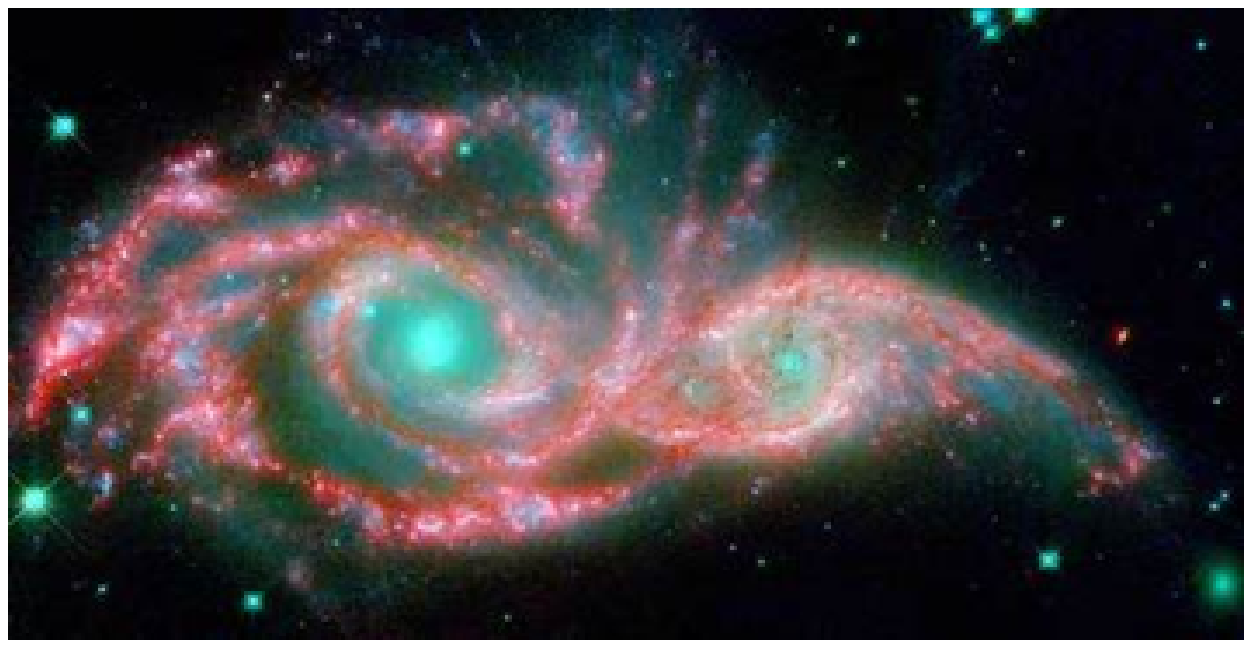}
\caption{Two examples of starbursting interactions/mergers (the Antennae and the NGC2207+IC2163 pair), where a large part of the starburst activity is not limited to the central kpc regions, but also takes places in super star clusters and giant HII regions throughout the disturbed disks, and in overlap regions between the interacting galaxies.}\label{fb6}
\end{figure}

{\bf \noindent The spatial extent of merger-induced starbursts}
The usual explanation of merger-induced starbursts accounts only for nuclear starbursts. While these are relatively frequent among mergers, they do not necessarily dominate interaction-triggered star formation, and spatially extended are actually common (see above, and two examples of Figure~\ref{fb6}). Quantitative comparisons of the extent of star formation in observations to that predicted by "standard" models have shown a significant disagreement \citet{barnes04,chien-barnes10}. These authors highlighted the failure of standard models to account for the extended morphology of (at least some) merger-induced starbursts. They also suggested that a sub-grid model of shock-induced star formation may better account for the spatial extent of merger-induced activity (see also \citealt {saitoh}).

\subsection{Star formation in mergers: effect of the turbulent, cloudy ISM}

The usually invoked mechanism of gas inflows for merger-induced star formation, as reproduced in low-resolution simulations, certainly takes place in real mergers -- signs of nuclear merger-driven starbursts are a plenty. But it seems impossible to explain the typical intensity of merger-induced starbursts and their often relatively larger spatial extend, based on these "standard" models.

Traditional SPH simulations model a relatively warm gas for the ISM, because the limited spatial resolution translates into a minimal temperature under which gas cooling should not be modeled (it would generate artificial instabilities, \citealt{truelove}). Modeling gas cooling substantially below $10^4$~K requires "hydrodynamic resolutions"\footnote{i.e., the average smoothing length in SPH codes, or the smallest cell size in AMR codes} better than 100~pc. Cooling down to 100~K and below can be modeled only at resolutions of a few pc. The vast majority of existing merger simulations hence have a sound speed of at least 10~km/s and cannot explicitly treat the supersonic turbulence that characterizes most of the mass in the real ISM \citep{burkert-ism}. Turbulent speeds in nearby disk galaxies are of  5-10~km/s, for sound speed of the order of 1-2 km/s in molecular clouds, i.e. turbulent Mach numbers up to a few. These are even higher in high-redshift disks \citep{FS09}, and in mergers (references above), but not necessarily for the same reason.

Increased ISM turbulence in galaxy mergers is thus absent from the modeling used in most hydrodynamic simulations to date. Some particle-based models have nevertheless been successful in reproducing these increased gas dispersions \citep{E93,struck97,bournaud08}, indicating that it is a consequence of the tidal interaction which triggers non-circular motions, rather than a consequence of starbursts and feedback. It should then arise spontaneously in hydrodynamic models, if these a capable of modeling gas below $10^{3-4}$K.

\paragraph{High-resolution simulations with multiphase ISM dynamics}

Adaptive Mesh Refinement (AMR) codes allow hydrodynamic calculations to be performed at very high resolution on adaptive-resolution grids. The resolution is not high everywhere, but the general philosophy is to keep the Jeans length permanently resolved until the smallest cell size is reached. That is, the critical process in the collapse of dense star-forming clouds, namely the Jeans instability (or Toomre instability in a rotating disk) is constantly resolved up to a typical scale given by the smallest cell size, or a small multiple of it. AMR simulations of whole galaxies have recently reached resolutions of a few pc for disk galaxies \citep[e.g.,]{agertz}, and even 0.8~pc lastly \citep{bournaud-lmc}. Such techniques have been first employed to model ISM dynamics and star formation in galaxy mergers by \citet{kim,teyssier10}. The main physical improvement is not necessarily directly the higher spatial resolution, but the possibility to model gas cooling down to low temperatures (below 300~K) while still resolving the typical Jeans length of gravitational instabilities. In theory, SPH techniques could also allow low temperature floor at high spatial resolution, but in practice this may be more expansive if not prohibitive, and has not been done for mergers and star formation (but see \citealt{hopkins11} but high-resolution, low-temperature floor SPH disk models).
 
 \smallskip
 
We here illustrate the results of such simulations using a sample of AMR simulations of 1:1 mergers of Milky Way-type spiral galaxies, performed with a resolution of 4.5~pc and a barotropic cooling model down to $\sim$50~K, technically similar to the isolated disk simulation described in \citep{bournaud-lmc}. Star formation takes place above a fixed density threshold and is modeled with a local Schmidt law, i.e. the local star formation rate density in each grid cell is $\rho_{\mathrm{SFR}} = \epsilon_{ff} \rho / t_{ff} \propto \rho^{1.5}$ where $\rho$ is the local gas density and $t_{ff}$ the gravitational free-fall time, and the efficiency $\epsilon_{ff}$ is a fixed parameter. Supernova feedback is included. Further details and results for whole sample will be presented in Powell et al. (in preparation). An individual model of this type (but at slightly lower resolution and without feedback), matching the morphology and star formation properties of the Antennae galaxies, was presented in \citealt{teyssier10}. More details on the simulations can also be found in \citep{bournaud-iau271}.
  \smallskip

In the models, the pre-merger isolated spiral galaxies spontaneously develop ISM turbulence at a about 10~km~s$^{-1}$ under the effect of gravitational instabilities (and/or feedback), and most star formation takes place in dense complexes of dense gas along spiral arms. In some sense, the large-scale star formation process is not entirely sub-grid anymore in these simulations, as the first steps of star formation, namely are the development of ISM turbulence and the formation of dense molecular gas clouds in this turbulent ISM, are explicitly captured -- the subsequent steps of star formation at smaller scales, inside the densest parts of these cold clouds, remain sub-grid.

\begin{figure}
\centering \includegraphics[width=6.5cm]{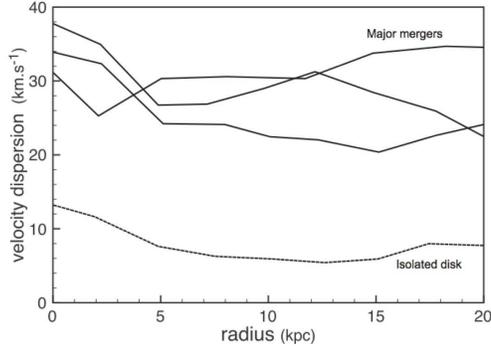}
\caption{Radial profiles of the gas velocity dispersion (turbulent speed) in an isolated galaxy simulation, and three major mergers between two disks of the same initial type.}\label{fb7}
\end{figure}

A merger simulation is shown in Figure~\ref{fb9}. The mass-weighted average of the gas velocity dispersion reaches $30-40$~km/s (Fig.~\ref{fb7}). This strong turbulence is consistent with the observations reviewed above. It induces numerous local shocks that increase the local gas density, which in turn triggers the collapse of gas into cold clouds. Also, gas clouds become more massive and denser than in the pre-merger spiral galaxies. The fraction of gas that is dense-enough and cold-enough to form star increases, and the timescale for star formation in these dense gas entities (the gravitational free-fall time) becomes shorter. As a result, the total SFR becomes several times higher than it was in the pre-interaction pair of galaxies. The standard process of merger-induced gas inflow towards the central kpc or so is also present, but the timescale is substantially longer, so this process dominates the triggering the star formation by enhancing the global gas density only in the late stages of the merger.

An example of star formation distribution is shown in Figure~\ref{fb9}. Two consequences of modeling a cold turbulent ISM in merging galaxies are: (1) the peak intensity of the starburst can become stronger (as shown by Teyssier et al. 2010 for a model of the Antennae) although this is not a systematic effect, and (2) the spatial extent of star formation in the starbursting phase is larger. The radius containing 50\% of the star formation rate (half-SFR radius) can more than double. This is because increased ISM turbulence is present throughout the systems (see Fig.~\ref{fb7}): this triggers, through locally convergent flows and shocks, the formation/collapse of efficiently star-forming clouds even at relatively large distances from the nuclei. At least quantitatively, these results put the models in better agreement with observations. In these models, the increased ISM turbulence is also obtained without feedback, showing that it is not a consequence of the starbursting activity, but is rather driven by the tidal forces in the interaction.

\begin{figure}
\centering \includegraphics[width=9.5cm]{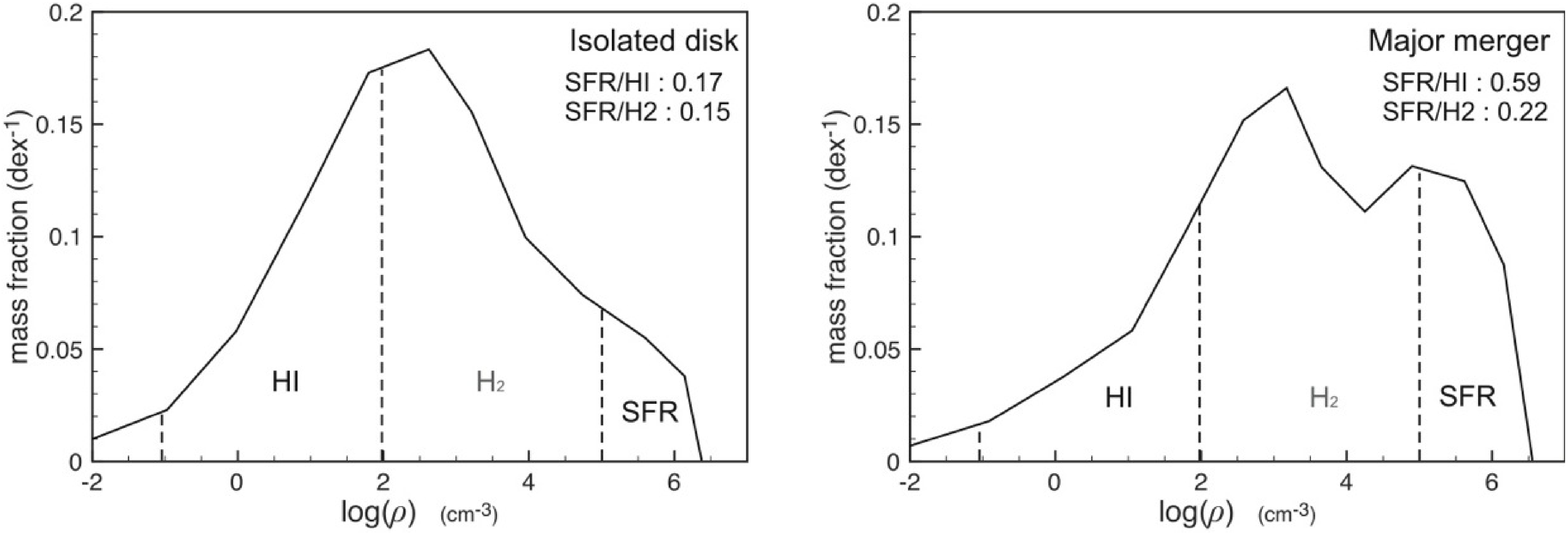}
\caption{Density PDFs of an isolated disk (left) and a starbursting major merger (right), in AMR simulations. The excess of very dense, star-forming gas results from the non log-normal shape of the PDF, induced by the rapidly increased turbulence and numerous local shocks compressing the gas, more than by a global shift of the initial PDF.}\label{fb8}
\end{figure}

\begin{figure}
\centering \includegraphics[width=9cm]{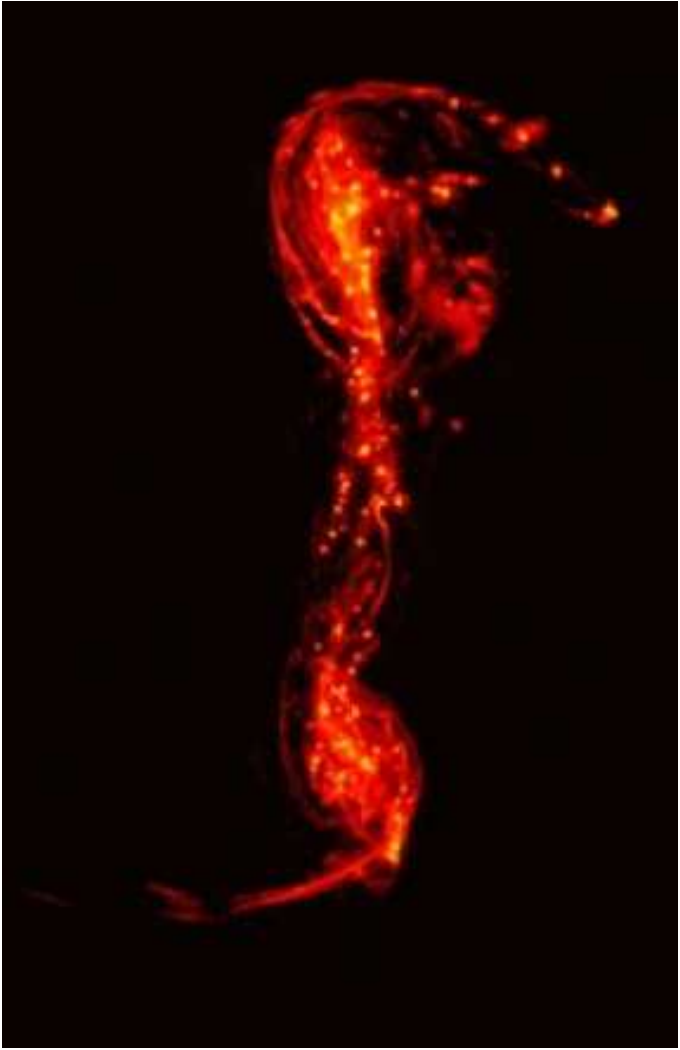}
\caption{View of the very young stars (proxy for the instantaneous star formation rate) in a major merger model from Teyssier et al. (2010).}\label{fb9}
\end{figure}

\begin{figure}
\centering \includegraphics[width=7.5cm]{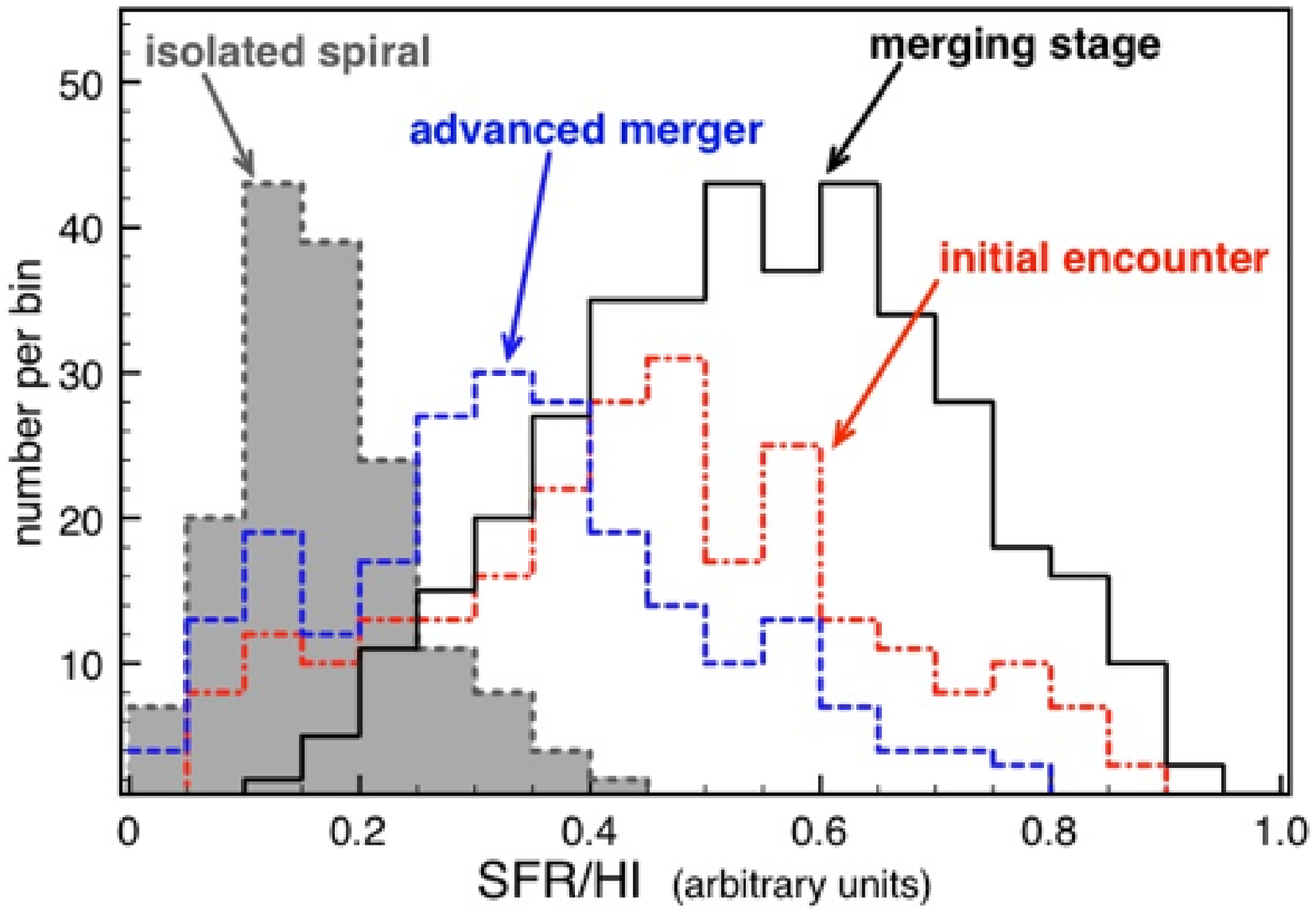}
\caption{Statistical distribution of the SFR/HI ratio (measured as the ratio of gas denser than $10^5$~cm$^{-3}$ as a proxy for SFR and gas less dense than $10^2$~cm$^{-3}$ as a proxy for HI (as illustrated on fig.~\ref{fb8}), for isolated spirals and mergers at different stages. This quantified the excess of dense gas illustrated earlier by density PDFs.}\label{fb10}
\end{figure}

\subsection{From triggered turbulence to dense gas excess and starbursts}

Modern models are thus capable of resolving ISM turbulence and SF clouds in disk galaxies and starbursting mergers. ISM turbulence is self-regulated in isolated disk galaxies, and is significantly triggered in merging systems.
  
Our simulated of disk galaxies have quasi log-normal PDFs (see example on Fig.~\ref{fb8}). The spatial resolution limit converts into a density limit at which the density PDF is truncated, and which corresponds to the smallest/densest entities that can be resolved. The maximal density resolution in the models presented here is around $10^{6}$~cm$^{-3}$, but parsec-scale resolution could capture even higher densities (as in Bournaud et al. 2010).
 
 \medskip
  
Representative density PDFs are shown for merger models on Figure~\ref{fb8} (see also Fig.~\ref{fb10}) for a moderately starbursting merger (SFR increased by a factor $\sim 3-4$ compared to the two pre-merger galaxies taken separately), and a stronger merger-induced starburst (factor $\sim 10$).  The PDFs of merging galaxies have a larger half-maximum width than those isolated disk galaxies, as expected from the higher turbulent speeds that result from the tidal interaction. Also, these PDFs are not  log-normal, as would be the case for isolated disks with a self-regulated turbulence cascade in a steady state. Rapid perturbations of the velocity field in mergers result in a substantial excess of dense gas in the density PDF. Such density PDFs naturally imply high SFRs, independently from the local SFR prescription, since the fraction of efficiently star-forming gas is high.

High fractions of dense gas in mergers were already proposed by \citep{juneau09}, based on detailed post-processing of merger simulations aimed at re-constructing dense molecular gas phases not resolved in these simulations \citep{narayanan}. Here we obtain a qualitatively similar conclusion using simulations that explicitly resolve turbulent motions, local shocks and small-scale instabilities in cold ISM phases. This excess of dense gas should have signatures in molecular line ratios. If, in a rough approach, we assume that low-$J$ CO lines are excited for densities of 100~cm$^{-3}$ and above, and HCN lines for densities of $10^4$~cm$^{-3}$ and above, then the HCN/CO line ratios could be $\sim$ 5-20 times higher in the starbursting phases of major mergers. Simulations with an somewhat higher resolution would actually be desirable to accurately quantify the emission of dense molecular tracers, and opacity effects may also affect the observed HCN/CO ratios.

\subsection{Implications for star formation "laws"}

The interpretation of merger-induced starbursts proposed from our high-resolution models is that it is not just a global gas inflow that increases the average gas density and increases the SFR, but also that there are strong non-circular motions, high turbulent velocity dispersions, causing many small-scale convergent flows and local shocks, that in turn initiate the collapse of dense star-forming clouds with high Jeans masses. The former ''standard'' process does take place, but the later can be equally important especially in the early phases of mergers.

We here note $\Sigma_{\mathrm gas}$ the average gas surface density of a galaxy. This is the quantity that observers would typically derive from the total gas mass and half-light radius, or similar quantities. The second mechanism above is a way to increase the SFR of a system, and its SFR surface density $\Sigma_{\mathrm SFR}$, without necessarily increasing its average $\Sigma_{\mathrm gas}$. Actually in our merger models $\Sigma_{\mathrm gas}$ does increase (as there are global merger-induced gas inflows), but $\Sigma_{\mathrm SFR}$ increases in larger proportions (as the starburst is not just from the global merger-induced inflow but also from the exacerbated fragmentation of high-dispersion gas). Going back to the density PDFs shown previously, one can note that the fraction of very dense gas (say, in the $\sim 10^{4-6}$~cm$^{-3}$ range) can increase by a factor of 10--20 in mergers while the average surface density $\Sigma_{\mathrm gas}$ increases by a factor 3--5 (see also on Figure~4). As a consequence, the $\Sigma_{\mathrm SFR}$ activity of these systems is unexpectedly high compared to their average surface density $\Sigma_{\mathrm gas}$.

Figure~4 shows the evolution of a system throughout a merger simulation in the  
$ \left( \Sigma_{\mathrm gas} ; \Sigma_{\mathrm SFR} \right)$ plane. While our pre-merger spiral galaxy models lie on the standard Kennicutt relation, starbursting mergers have high $\Sigma_{\mathrm SFR} / \Sigma_{\mathrm gas}$ ratios. This is in agreement with observational suggestions that quiescent disks and starbursting mergers do not follow the same scaling relations for star formation, but could actually display two different star formation "laws" \citep{daddi10b, genzel10}. The offset between the disk and merger sequences proposed by Daddi et al. (2010b) is quantitatively recovered in our simulations (Figure~4). Post-starburst, post-merger systems lie back on the quiescent sequence, or even somewhat below it: these systems contain some dense gas which is somewhat stabilized by the stellar spheroid. This is another example of a "morphological quenching" effect in early-type galaxies \citep{martig09}, and the location of our post-merger early-type galaxies in the $ \left( \Sigma_{\mathrm gas} ; \Sigma_{\mathrm SFR} \right)$ diagram {\em may} be consistent with observations of nearby ellipticals \citep{crocker10,saintonge2}, without having a substantial effect on the formation of the red sequence -- at least at low redshift \citep{fabello}, .

\begin{figure}
\centering \includegraphics[width=6cm]{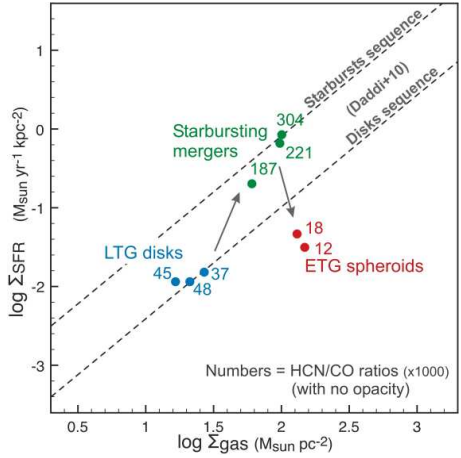}
\caption{Kennicutt diagram comparing the surface density of gas and star formation in isolated disk galaxies (LTGs) and mergers near the peak of their starburst activity, as well as post-merger elliptical-like galaxies (ETGs). The "disks" and "starbursts" sequences are from Daddi et al. (2010). The low star formation efficiency in ellipticals is studied in \citep{martig09}.}\label{fb11}
\end{figure}

The proposal that disks and mergers follow two different regimes of star formation by Daddi et al. (2010b) and Genzel et al. (2010) relies for a part (but not entirely) on the assumption that different CO luminosity-to-molecular gas mass conversion factors apply in quiescent disks and starbursting mergers. Interestingly, our simulations recover the two regimes of star formation without any assumption on such conversion factors since gas masses are directly known. But at the same time, excess of dense gas found in these merger models suggests that the excitation of CO lines would naturally be higher in mergers/starbursting phases (although this needs to be quantified in the models), which would mean that the assumption of different conversion factors by Daddi et al. and Genzel et al. could be physically justified. High molecular gas excitation in SubMillimeter Galaxies (SMGs, Tacconi et al. 1998) could then naturally result if these are starbursting major mergers with high gas surface densities and a clumpy turbulent ISM \citep[e.g.][]{narayanan, bournaud-iau271}. However models do not predict a bimodality. Two regimes of star formation appear when mergers are only selected at the peak of their starburst activity, but merging systems spend over half of their time below the "starburst regime" on the diagram shown in Figure~\ref{fb11}.

\subsection{Summary}

The triggering of star formation in merger largely results from tidal inflows driven by the interaction, which fuels a relatively concentrated or "nuclear" starburst. This is the main mechanism highlighted by resolution-limited simulations for two decades. Nevertheless it does not correctly account for observations of starbursting mergers: real mergers often have a nuclear starburst component, but also a relatively extended starburst activity, often taking place in massive star clusters. Recent simulations that resolve ISM turbulence and dense gas clouds show that the triggering of star formation in mergers is partly from gas inflows, but also from increased ISM turbulence, fragmentation into bigger/denser clouds, resulting in an excess of dense star forming gas compared to atomic gas or moderately-dense molecular gas. This can explain a "starburst regime" of star formation that does not follow the global scaling relations observed for isolated disk galaxies.


\section{Structure formation in mergers, Super Star Clusters and Globular Clusters}

\begin{figure}
\centering \includegraphics[width=6cm]{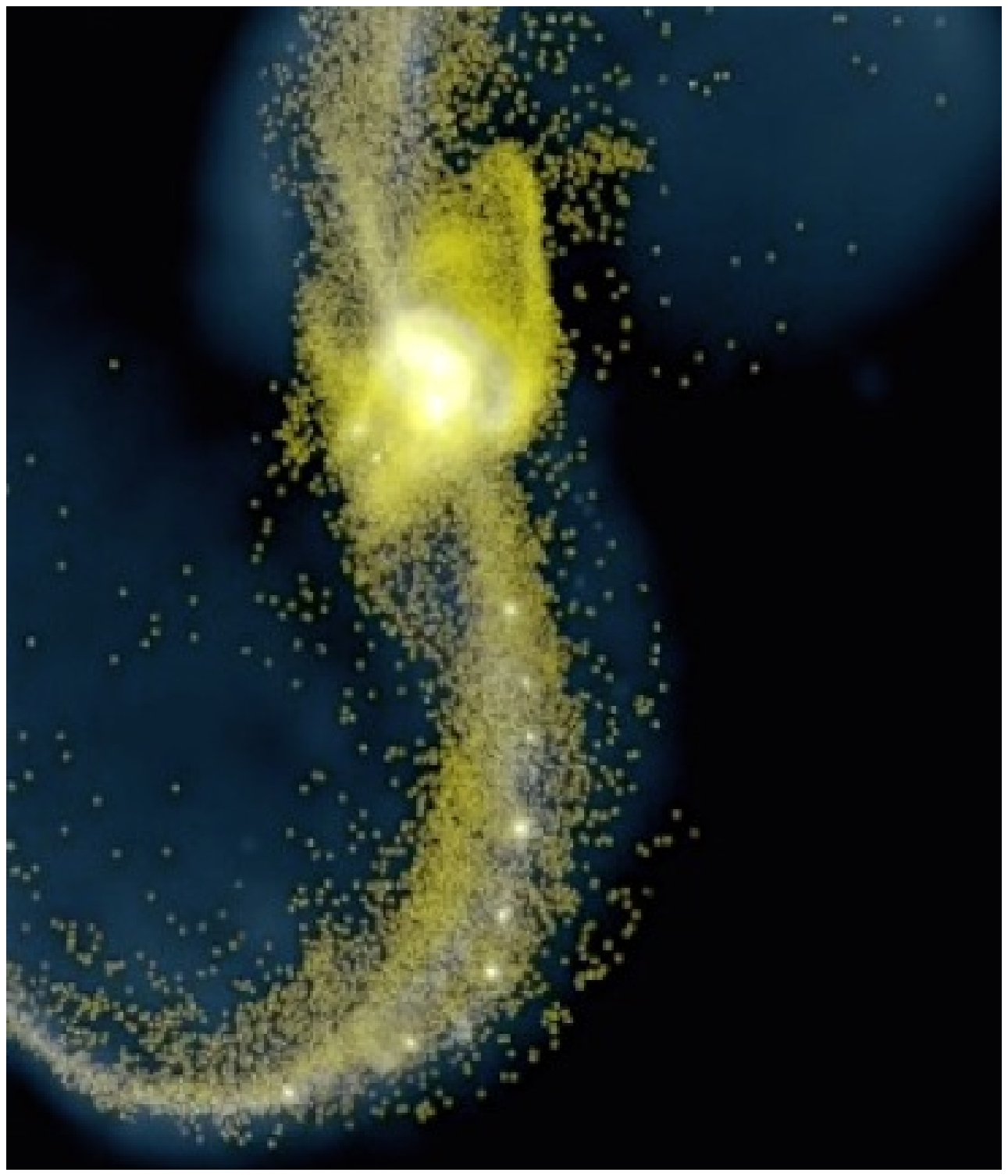}
\includegraphics[width=6cm]{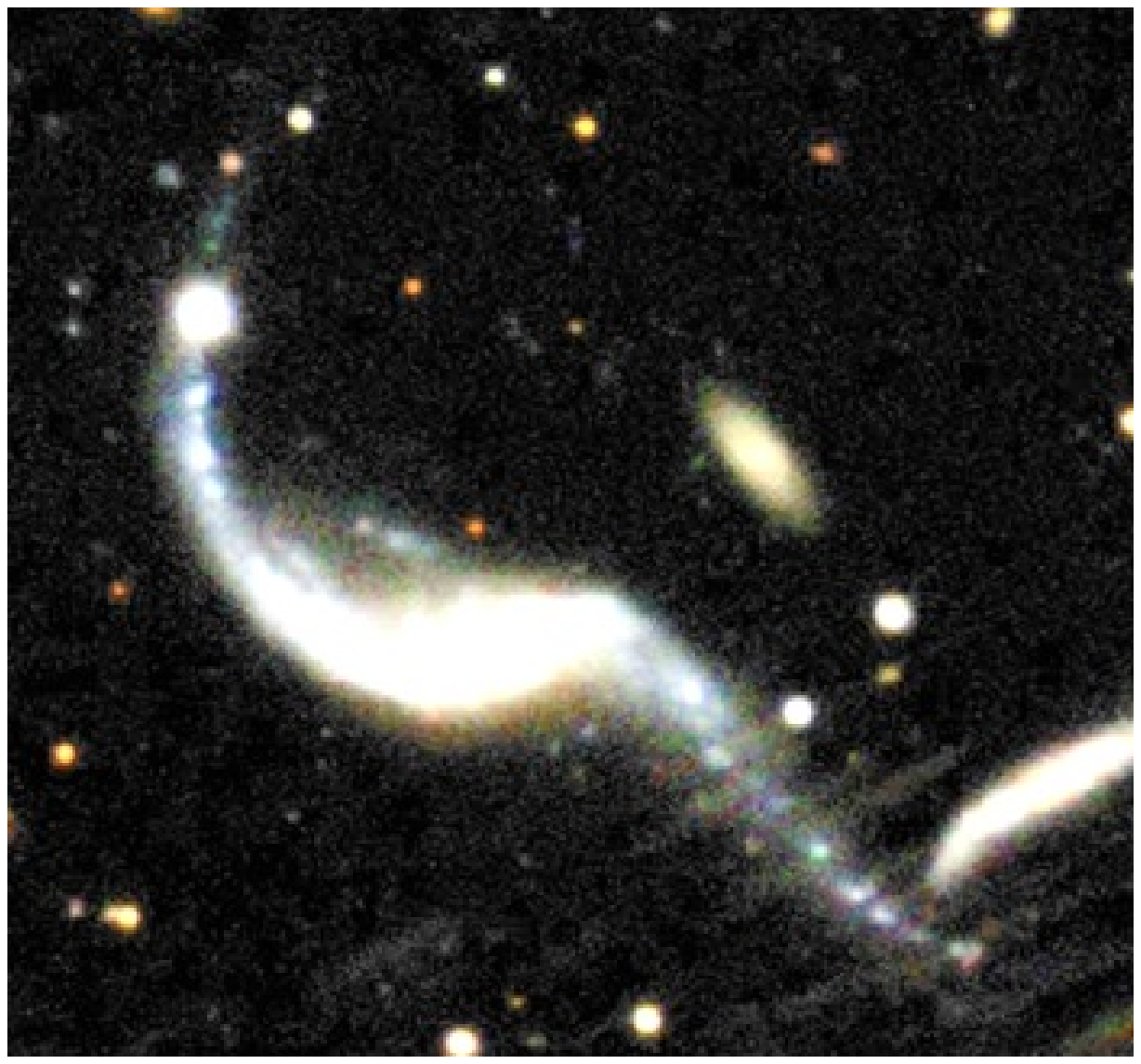}
\caption{Simulations (left, \citealt{wetzstein}) and observations (right) of massive star cluster formation in a "beads-on-a-string" mode, typical for Jeans instabilities in tidal tails.}\label{fb12}
\end{figure}

Galaxy interactions and mergers lead to bursts of star formation, but also play a major role in forming new stellar structures: rings, tails, shells, and more importantly massive star clusters that could be the progenitors of globular clusters and, maybe in some cases, of tidal dwarf galaxies and/or ultracompact dwarf galaxies. Examples of formation of massive/compact star clusters in the tidal tails of major mergers are shown on Fig.~\ref{fb12}. A detailed review of structure formation and star cluster formation/evolution in galaxy interactions and mergers can be found in \citet{bournaud-clusters}.


\section{Conclusion and implications for galaxy formation}

Galaxy interactions and mergers are a major driver of cosmic star formation, even if their contribution to the SF budget in the Universe compared to quiescent star formation in isolated disk galaxies is still unclear and debated. 

Gas compression in galaxy interactions and mergers results from two mechanisms: a global one (radial gas inflows towards the nucleus) and a local one (compression and fragmentation of the ISM in turbulent flows with a Mach number well above one). The latter mode was really probed only recently, although observations of triggered ISM turbulence in interactions have existed for more than a decade. Numerical simulations are hence barely starting to model merger-induced starbursts in a realistic way, and cosmological models probably do not account for these events in a realistic way, which will be a crucial aspect to understand the global star formation history of galaxies. 

Nevertheless, recent accurate observations of ISM properties and star formation in mergers can be accounted by theoretical models. These include a "double law" of star formation for mergers versus isolated disks, unveiling a "starburst mode" and a "quiescent mode", as well as an excess of dense molecular gas in the ISM of starbursting mergers. This probably shows that our theoretical understanding of the driving processes has improved and that modern simulations can model these in a realistic way. Major issues nevertheless remain, such as the role of stellar feedback processes, the triggering of gaseous outflows, the influence of gas accretion by central supermassive black holes in mergers, the mergers of such black holes into bigger and bigger black holes but with gravitational recoil effects, and the energetic feedback from these black holes.




\begin{thebibliography}{99}

\bibitem[Agertz et al.(2009)]{agertz} Agertz, O., et al. 2009, MNRAS, 392, 294

\bibitem[Appleton 
\& Struck-Marcell(1996)]{appleton} Appleton, P.~N., \& Struck-Marcell, C.\ 1996, Fundamentals of Cosmic Physics, 16, 111 


\bibitem[Athanassoula(1992)]{athanassoula} Athanassoula, E.\ 1992, 
MNRAS, 259, 345 

\bibitem[Barnes \& Hernquist(1991)]{BH92} Barnes, J.~E., \& Hernquist, L.~E.\ 1991, ApJL, 370, L65 

\bibitem[Barnes(2004)]{barnes04} Barnes, J.~E.\ 2004, MNRAS, 350, 798 

\bibitem[Bell et al.(2006)]{bell} Bell, E.~F., Phleps, S., 
Somerville, R.~S., Wolf, C., Borch, A., \& Meisenheimer, K.\ 2006, ApJ, 652, 270 

\bibitem[Block et al.(2010)]{block} Block, D.~L., Puerari, 
I., Elmegreen, B.~G., \& Bournaud, F.\ 2010, Apjl, 718, L1 

\bibitem[Bournaud et al.(2004)]{bournaud-fp} Bournaud, F., Duc, P.-A., Amram, P., Combes, F., \& Gach, J.-L.\ 2004, A\&A, 425, 813 

\bibitem[Bournaud et al.(2007)]{B07} Bournaud, F., et al.\ 
2007, Science, 316, 1166 

\bibitem[Bournaud et al.(2008)]{bournaud08} Bournaud, F., Duc, P.-A., \& Emsellem, E.\ 2008, MNRAS, 389, L8 

\bibitem[Bournaud(2010a)]{bournaud-clusters} Bournaud, F.\ 2010a, Galaxy Wars: Stellar Populations and Star Formation in Interacting Galaxies, 423, 177 

\bibitem[Bournaud(2010b)]{bournaud-iau271} Bournaud, F., Powell, 
L.~C., Chapon, D., \& Teyssier, R.\ 2010d, in IAU Symposium 271, N. Brummell, A. S. Brun, M. S. Miesch \& Y. Ponty Eds. ; arXiv:1012.5227 


\bibitem[Bournaud et al.(2010c)]{bournaud-lmc} Bournaud, F., Elmegreen, B.G., Teyssier, R., Block, D.L., \& Puerari, I. 2010c, MNRAS 409, 1088


\bibitem[Burkert(2006)]{burkert-ism} Burkert, A.\ 2006, Comptes Rendus Physique, 7, 433

\bibitem[Chien \& Barnes(2010)]{chien-barnes10} Chien, L.-H., \& Barnes, J.~E.\ 2010, MNRAS, 407, 43 


\bibitem[Combes \& Gerin(1985)]{CG85} Combes, F., \& Gerin, M.\ 1985, A\&A, 150, 327 

\bibitem[Conselice et al.(2003)]{conselice-cas} Conselice, C.~J., 
Bershady, M.~A., Dickinson, M., \& Papovich, C.\ 2003, AJ, 126, 1183 


\bibitem[Cox et al.(2008)]{cox-ratios} Cox, T.~J., Jonsson, P., Somerville, R.~S., Primack, J.~R., \& Dekel, A.\ 2008, MNRAS, 384, 386 

\bibitem[Crocker et al.(2011)]{crocker10} Crocker, A.~F., Bureau, M., Young, L.~M., \& Combes, F.\ 2011, MNRAS, 410, 1197 

\bibitem[Cullen et al.(2006)]{cullen06} Cullen, H., Alexander, P., \& Clemens, M.\ 2006, MNRAS, 366, 49 

\bibitem[Daddi et al.(2010a)]{daddi10a} Daddi, E., et al.\ 2010a, ApJ, 713, 686

\bibitem[Daddi et al.(2010b)]{daddi10b} Daddi, E., et al.\ 2010b, ApJL, 714, L118 

\bibitem[Di Matteo et al.(2007)]{dimatteo07} Di Matteo, P., Combes, F., Melchior, A.-L., \& Semelin, B.\ 2007, A\&A, 468, 61 

\bibitem[Di Matteo et al.(2008)]{dimatteo08} Di Matteo, P., Bournaud, F., Martig, M., Combes, F., Melchior, A.-L., \& Semelin, B.\ 2008, A\&A, 492, 31 


\bibitem[Dobbs et al.(2010)]{dobbs51} Dobbs, C.~L., Theis, C., 
Pringle, J.~E., \& Bate, M.~R.\ 2010, MNRAS, 403, 625 


\bibitem[Duc et al.(1997)]{duc97} Duc, P.-A., Brinks, E., Wink, J.~E., \& Mirabel, I.~F.\ 1997, A\&A, 326, 537 

\bibitem[Dutta et al.(2009)]{dutta} Dutta, P., Begum, A., Bharadwaj, S., \& Chengalur, J.~N.\ 2009, MNRAS, 397, L60 

\bibitem[Elbaz \& Cesarsky(2003)]{elbaz-cesarsky} Elbaz, D., \& Cesarsky, C.~J.\ 2003, Science, 300, 270 


\bibitem[Elmegreen \& Lada(1977)]{elm-lada} Elmegreen, B.~G., \& Lada, C.~J.\ 1977, ApJ, 214, 725 


\bibitem[Elmegreen \& Thomasson(1993)]{ET93} Elmegreen, B.~G., \& Thomasson, M.\ 1993, A\&A, 272, 37 

\bibitem[Elmegreen et al.(1993)]{E93} Elmegreen, B.~G., Kaufman, M., \& Thomasson, M.\ 1993, ApJ, 412, 90

\bibitem[Elmegreen et al.(1995)]{E95} Elmegreen, D.~M., Kaufman, M., Brinks, E., Elmegreen, B.~G., \& Sundin, M.\ 1995, ApJ, 453, 100


\bibitem[Elmegreen(2002)]{E02} Elmegreen, B.~G.\ 2002, ApJ, 577, 206 

\bibitem[Elmegreen et al.(2006)]{ngcic-spitzer-e06} Elmegreen, D.~M., 
Elmegreen, B.~G., Kaufman, M., Sheth, K., Struck, C., Thomasson, M., \& Brinks, E.\ 2006, ApJ, 642, 158 

\bibitem[Fabello et al.(2011)]{fabello} Fabello, S., Catinella, B., Giovanelli, R., Kauffmann, G., Haynes, M.~P., Heckman, T.~M., \& Schiminovich, D.\ 2011, MNRAS, 411, 993 

\bibitem[F{\"o}rster Schreiber et al.(2009)]{FS09} F{\"o}rster Schreiber, N.~M., et al.\ 2009, ApJ, 706, 1364 

\bibitem[Genzel et al.(2010)]{genzel10} Genzel, R., et al.\ 2010, MNRAS, 407, 2091 

\bibitem[Hancock et al.(2009)]{hancock} Hancock, M., Smith, B.~J., Struck, C., Giroux, M.~L., \& Hurlock, S.\ 2009, AJ, 137, 4643 

\bibitem[Hopkins et al.(2011)]{hopkins11} Hopkins, P.~F., Quataert, E., \& Murray, N.\ 2011, arXiv:1101.4940 

\bibitem[Irwin(1994)]{irwin} Irwin, J.~A.\ 1994, ApJ, 429, 618 

\bibitem[Jogee et al.(2009)]{jogee} Jogee, S., et al.\ 2009, ApJ, 697, 1971 

\bibitem[Juneau et al.(2009)]{juneau09} Juneau, S., Narayanan, D.~T., Moustakas, J., Shirley, Y.~L., Bussmann, R.~S., Kennicutt, R.~C., \& Vanden Bout, P.~A.\ 2009, ApJ, 707, 1217 

\bibitem[Keel et al.(1985)]{kennicutt87} Keel, W.~C., Kennicutt, 
R.~C., Jr., Hummel, E., \& van der Hulst, J.~M.\ 1985, AJ, 90, 708 

\bibitem[Kim et al.(2009)]{kim} Kim, J.-h., Wise, J.~H., \& Abel, T.\ 2009, ApJL, 694, L123 

\bibitem[Krumholz \& Thompson(2007)]{K07} Krumholz, M.~R., \& Thompson, T.~A.\ 2007, ApJ, 669, 289 

\bibitem[Lawrence et al.(1989)]{lawrence89} Lawrence, A., Rowan-Robinson, M., Leech, K., Jones, D.~H.~P., \& Wall, J.~V.\ 1989, MNRAS, 240, 329 

\bibitem[Lombardi et al.(2010)]{alves} Lombardi, M., Lada, C.~J., \& Alves, J.\ 2010, A\&A, 512, A67 

\bibitem[Lotz et al.(2008)]{lotz-gm20} Lotz, J.~M., Jonsson, P., 
Cox, T.~J., \& Primack, J.~R.\ 2008, MNRAS, 391, 1137 

\bibitem[Martig \& Bournaud(2008)]{MB08} Martig, M., \& Bournaud, F.\ 2008, MNRAS, 385, L38 

\bibitem[Martig et al.(2009)]{martig09} Martig, M., Bournaud, F., Teyssier, R., \& Dekel, A.\ 2009, ApJ, 707, 250

\bibitem[Mihos \& Hernquist(1996)]{mihos} Mihos, J.~C., \& Hernquist, L.\ 1996, ApJ, 464, 641 

\bibitem[Narayanan et al.(2010)]{narayanan} Narayanan, D., et al. 2010, MNRAS, 401, 1613

\bibitem[Petersen et al.(2007)]{petersen} Petersen, M.~R., 
Julien, K., \& Stewart, G.~R.\ 2007, ApJ, 658, 1236 

\bibitem[Renaud et al.(2008)]{renaud} Renaud, F., Boily, 
C.~M., Fleck, J.-J., Naab, T., \& Theis, C.\ 2008, MNRAS, 391, L98 

\bibitem[Robaina et al.(2009)]{robaina09} Robaina, A.~R., et al.\ 2009, ApJ, 704, 324 

\bibitem[Saintonge et al.(2011)]{saintonge2} Saintonge, A., et al.\ 2011, arXiv:1104.0019 

\bibitem[Saitoh et al.(2009)]{saitoh} Saitoh, T.~R. et al. 2009, PASJ, 61, 481 

\bibitem[Sanders et al.(1988)]{sanders88} Sanders, D.~B., et al. 1988, ApJ, 325, 74 

\bibitem[Shapiro et al.(2008)]{shapiro08} Shapiro, K.~L., et al.\ 
2008, ApJ, 682, 231 

\bibitem[Smith et al.(2008)]{smith} Smith, B.~J., et al.\ 2008, AJ, 135, 2406 

\bibitem[Soifer et al.(1984)]{soifer84} Soifer, B.~T., et al.\ 
1984, ApJL, 278, L71 

\bibitem[Struck(1997)]{struck97} Struck, C.\ 1997, ApJS, 113, 269 

\bibitem[Tacconi et al.(2010)]{tacconi10} Tacconi, L.~J., et al.\ 
2010, Nature, 463, 781 

\bibitem[Teyssier et al.(2010)]{teyssier10} Teyssier, R., Chapon, D., \& Bournaud, F.\ 2010, ApJL 720, 149

\bibitem[Toomre(1964)]{toomre64} Toomre, A.\ 1964, ApJ, 139, 
1217 

\bibitem[Toomre \& Toomre(1972)]{TT72} Toomre, A., \& Toomre, J.\ 1972, ApJ, 178, 623 

\bibitem[Toomre \& Toomre(1977)]{toomre77} Toomre, A., \& Toomre, J.\ 1977, The New Astronomy and Space Science Reader, 271 

\bibitem[Toomre(1981)]{toomre81} Toomre, A.\ 1981, Structure and 
Evolution of Normal Galaxies, 111 

\bibitem[Truelove et al.(1997)]{truelove} Truelove, J.~K., et al.\ 1997, ApJl, 489, L179

\bibitem[van den Bergh(2002)]{vandenbergh} van den Bergh, S.\ 2002, AJ, 124, 786 

\bibitem[Wada et al.(2002)]{wada02} Wada, K., Meurer, G., \& Norman, C.A. 2002, ApJ, 577, 197

\bibitem[Wang et al.(2004)]{wang04} Wang, Z., et al.\ 2004, ApJS, 154, 193 

\bibitem[Wetzstein et al.(2007)]{wetzstein} Wetzstein, M., Naab, T., \& Burkert, A.\ 2007, MNRAS, 375, 805 


\end{thebibliography}
\end{document}